\documentclass[preprint,superscriptaddress]{revtex4}
\usepackage{graphics,epsfig}
\usepackage{epstopdf}
\usepackage{graphicx}
\usepackage{dcolumn}
\usepackage{amsmath}
\newcommand{\be}{\begin{equation}}
\newcommand{\ee}{\end{equation}}
\newcommand{\bea}{\begin{eqnarray}}
\newcommand{\eea}{\end{eqnarray}}
\newcommand{\nn}{\nonumber}
\def\be{\begin{equation}}
\def\ee{\end{equation}}
\def\ba{\begin{eqnarray}}
\def\ea{\end{eqnarray}}
\begin{document}
\title{On Thermodynamics and Statistical Entropy of Bardeen Black Hole}

\author{Dharm Veer Singh}
\email{veerdsingh@gmail.com}
\affiliation{Department of Physics, Institute of Applied Science and Humanities, G. L. A. University, Mathura, 281406 India}
\author{Sanjay Siwach}
\email{sksiwach@hotmail.com}
\affiliation{Department of Physics, Institute of Science, Banaras Hindu university, Varanasi 221005, India}

\begin{abstract}
\noindent We discuss the first law of thermodynamics of Bardeen black hole. The presence of magnetic monopole charge modifies the energy and we define the temperature of the system accordingly. We introduce a lower cut-off to the spacetime by introducing a spherical surface, which can be identified with the event horizon. The corresponding temperature is identified with Hawking temperature. The entropy is assumed to obey the area law and we confirm this by calculating the statistical entropy by using brick wall model.     
\end{abstract} 

\maketitle


The black holes obey the laws of thermodynamics similar to a thermodynamics system. The temperature of the black hole is related to the surface gravity and the entropy obeys the area law \cite{JD}
\be
S_{BH}=\frac{k_B}{4}\frac{A}{l^2_p},
\ee
where $l_P=\sqrt{\hbar G/c^3}$ is the Planck length, $k_B$ is a Boltman constant, $G$ is gravitational constant and $c$ is speed of light. This area-entropy law is valid for all black holes but subject to quantum gravity corrections. 

However, there is a class of regular black holes where, it seems to be violated \cite {fr1,dvs18,ads,Maluf:2018lyu,Ansoldi:2008jw,javed}. The  Bardeen black hole \cite {Bardeen:1968} belong to this class. Bardeen black hole was given a new meaning by Ayon-Beato {\it et al} using the nonlinear electromagnetic source \cite{ABG99}. In fact the magnetic monopole charge seem to prevent the occurrence of the singularity. Since then many black holes solutions have been constructed by introducing nonlinear electromagnetic source \cite{Xiang,Singh:2017qur,dym1,hamid18,singh18}. The problem of area law violation for these black holes arise due to the inconsistency with first law of thermodynamics. The problem is resolved if one uses the modified form of first law of thermodynamics \cite{ma14}. 

Here, we revisit the problem and suggest to consider the energy of the system as if there is some lower cut-off to the system introduced by a spherical surface. In the case of black hole this surface will be identified with the location of event horizon. The energy of the system consist of both gravitational energy due to black hole mass as well as contribution due to stress tensor of non-linear electrodynamics and depends on the location of the surface. The temperature of the system is identified with Hawking temperature if the lower cut-off is taken at the event horizon. 

The correct use of the first law of black hole thermodynamics, which includes the parameters of non-linear electromagntic Lagrangian  precisely reproduces the modification of first law mentioned above \cite{rashid,breton,zhang,gulin}. The entropy is assumed to obey the area law \cite{wald93} and this is confirmed by calculating the statistical  entropy of Bardeen black hole using the brick wall model \cite{hooft,li,aghosh,eune,sarkar,kghosh,solodukhin1} (Appendix).  

Let us consider  the  action of Einstein gravity in the presence of nonlinear electromagnetic field, 
\be
S=\frac{1}{16\pi}\int d^4x \sqrt{-g}\left[R-4{\cal L}(F)\right]. \label{action}
\ee

Here $R$ is Ricci scalar and ${\cal L}(F)$ is the Lagrangian of nonlinear electromagnetic field, given by \cite{ABG99},
\be
{\cal L}(F)=\frac{3}{2 s\, g^2}\left(\frac{\sqrt{2g^2F}}{1+\sqrt{2g^2F}}\right)^{\frac{5}{2}},
\ee
where $F=F_{\mu\nu}F^{\mu\nu}$, $F_{\mu\nu}$ is the electromagnetic field tensor, g is the magnetic monopole charge. The parameter $s$ is defined as $s = g/2M$, where $M$ will be identified as black hole mass.  

The above action admits regular spherical symmetric solutions with line element,
\be
ds^2=f{(r)}dt^2- \frac{1}{f(r)}dr^2 - r^2 d\Omega^2,
\label{metric}
\ee
and the Maxwell field is given by, $F_{\theta\phi}=-F_{\phi\theta}=g\sin \theta$ \cite{ABG99,fr1}.
The function $f(r)$ has the form,
\be
f(r)=1-\frac{2Mr^2}{(r^2+g^2)^{3/2}}.
\label{bar1}
\ee 

This solution is known as Bardeen spacetime \cite{Bardeen:1968}. Bardeen spacetime admits regular black hole solution for a range of parameters. In the limit $g=0$, we get the Schwarzchild metric, which is singular. The singularity seems to be resolved by the monopole charge in the case of Bardeen black hole.

The location of the horizon is determined by the condition, $f(r_h)=0$, which gives,
\be
M=\frac{(r^2_h+g^2)^{3/2}}{2r^2_h}.
\ee

The Hawking temperature of the black hole is defined  to be proportional to the surface gravity  $T_H=\frac{\kappa}{2\pi}$ and is given by,     
\be 
T_H=\frac{1}{4\pi}f'(r_h)=\frac{r_h^2-2g^2}{4\pi r_h(r_h^2+g^2)}.
\label{temp21}
\ee

The entropy of the black hole can be obtained by using the first law of thermodynamics ,
\be
S=\int^{r_h}_{0}\frac{1}{T}\left(\frac{dM}{dr}\right)dr.
\ee

If we take $T=T_H$, the entropy is found as \cite{sharif}, 
\be
S={\pi}\left[(r_h^2-2g^2)\sqrt{1+\frac{g^2}{r_h^2}}+3g^2r_h\log(1+\sqrt{1+\frac{g^2}{r_h^2}})\right].
\ee
However, this entropy is not proprtional to the area of event horizon and the second law of thermodynamics seems to be violated. 

The authors \cite{ma14} suggested that one can recover the area law for black hole entropy $S=\pi\,r^2_h$, provided one uses the modified form of the first law of thermodynamics for the Bardeen black hole, 
\be
C_MdM=T\, dS+\psi_q \, dq ,
\label{modf}
\ee
where, $C_M =\frac{r^3_h}{(r^2_h+g^2)^\frac{3}{2}}$ and $\psi_q$ is the conjugate potential of the magnetic charge, $q=g$.

To have some understanding of the issue, let us consider a spherical surface of radius $r_s$. This will serve as lower cut-off to the system and can be identified with the event horizon in real practice. Let the energy (gravitational + monopole contribution) of the system with this cut-off be denoted by $m(M,r_s)$ and let us denote the temperature of the system by $T(r_s)$, such that, 
\be
dm(M,r_s) = T\, dS + \psi_q \, dq.
\ee

The energy function $m(M,r_s)$ receives the contribution from black hole mass as well as stress tensor, $T^0_0$ of non-linear electromagnetic field and is given by \cite{ma14},
\be
m(M,r_s)=M+4\pi\int_{r_s}^{\infty} r^2 T^0_0\;dr.
\ee

If we insist that area law of the black hole entropy is correct and compare with the form of first law (\ref{modf}), we immediately get the relation,
\bea
T_H=C_M T_\infty,
\eea
where $T_\infty$ is the temperature of the system if the lower cut-off is pushed to infinity. In real practice, lower cut-off is provided by the location of event horizon. The temperature $T_\infty$ is just a convenient temperature, if we naively use the energy of the system without stress tensor contribution i.e. $M$.

For Bardeen black hole, however, the temperature as measured by an asymptotic observer is $T_H$, which corresponds to modified energy of the system as $m(M,r_h)$ and  it suffices to consider a spherical surface with radius $r_s = r_h$.  Thus, the difficulty to obtain the area law of black hole entropy seems to be associated with the incorrect use of the first law of thermodynamics.  

Let us consider the general form of the first law of thermodynamics applied to a black hole in non-linear electrodynamics.
\be
dM=T\,dS + \psi_q\,dq + \psi_{a_i}\,da_i,
\ee
where $a_i$ are the parameters associated with nonlinear Lagrangian density, $\cal L$ and $\psi_{a_i}$ etc. denote the conjugate potentials.

The nonlinear Lagrangian density, $\cal L$ of the Bardeen solution contains two parameters, mass $M$ and the magnetic charge $q=g$ and the first law can be written as,
\be
dM=T\,dS + \psi_q\,dq + \phi_q\,dq + \phi_{M}\,dM,
\ee

which can be rewritten as,
\be
C_M dM=T\,dS + \Phi_q\,dq  
\ee

where $C_M = 1-\phi_{M}$ and $\Phi_q=\psi_q + \phi_q$.

The explicit form of $\phi_q$ and $\phi_{M}$ are given in the literature \cite{zhang,gulin} and we can find $C_M =\frac{r^3_h}{(r^2_h+g^2)^\frac{3}{2}}$ in agreement with the relation (\ref{modf}) obtained by \cite{ma14}.

Now the temperature $T=C_M\left(\frac{\partial M}{\partial S}\right)_{q}$, where $S=\pi r^2_h$ is the Bekenstein-Hawking entropy, is obtained as,
\be
T=\frac{(r_h^2-2g^2)}{4\pi r_h(r_h^2+g^2)},
\ee
which agrees with the Hawking temperature.

Similarly, $\Phi_q=C_M\left(\frac{\partial M}{\partial q}\right)_{S}$ is obtained as,
 \be
\Phi_q=  = \frac{3gr_h}{2(r^2_h+g^2)}.
\ee

One can easily very that various quantities obey the Smarr relation in 4-dimension,
\be
M=2TS+\Psi_q\, q + \phi_M \,M.
\ee

In summary, the correct use of first law of black hole thermodynamics enables us to recover the relation given by the authors \cite{ma14} and there is no need for the modification of first law or second of black hole thermodynamics. 

\newpage

\section{\bf{Appendix: Brick wall model}}

Let us consider the Kelin-Gordan field equation for the massive scalar field in the background (\ref{metric}) is
\bea
\frac{1}{\sqrt{-g}}\partial_\mu\left(\sqrt{-g}g^{\mu\nu}\partial_{\nu}\Phi\right)-{m^2_\phi}\Phi=0.
\eea
Let $\Phi=e^{-i\,E t}R(r)Y^m_l(\theta,\phi)$, then the equation for $R(r)$ becomes,
\ba
&&\frac{1}{f(r))}E^2R(r)+\frac{1}{r^2}\partial_r(r^2f(r)\partial_rR(r))\nn\\&&\qquad\qquad\qquad-\frac{l(l+1)}{r^2}R(r)-m^2R(r)=0.
\ea

The above equation can be solved in WKB approximation, and the density of states in the leading order can be obtained using brick wall method \cite{hooft},
\bea
\pi n_{E l}=\int_{r_h+\epsilon}^{L}k_{E\ell}\left(r\right)\;dr
\eea
where
\bea
k^2_{E,l}(r)=\frac{\left[E^2-f(r)\left(\frac{\ell\left(\ell+1\right)}{r^2}\right)+m^2_\phi\right]}{f^2\left(r\right)}
\eea
and the total number of the states with an energy less than $E$ is given by,
\bea
\pi N_E=\sum_{modes} n_{E,\ell}\simeq \int^{\ell_{max}}\left(2\ell+1\right)d\ell\int_{r_h+\epsilon}^{L}k_{E,\ell}\left(r\right)\;dr\nn.
\eea
The sum over $l$ can be performed and the free energy of the system is given by \cite{wint},
 \bea
F=-\frac{2}{3\pi}\int_{0}^{\infty}\frac{dE}{e^{\beta E}-1}\;\int_{r_h+\epsilon}^{L}\frac{r^2}{f^2(r)}\left[E^2-m^2_\phi f(r)\right]\;dr
\eea
The free energy is dominated by the modes near the event horizon and the upper limit just provide the vacuum energy proportional to the volume of the system. 

We may expand $f(r)$ close to event horizon using the Taylor series,
\bea
f(r)=f^\prime(r_h)\left(r-r_h\right)+f^{\prime\prime}(r_h)\left(r-r_h\right)^2+\ldots.
\eea

The entropy of the system is given by,
\bea
S = \beta^2\frac{\partial F}{\partial \beta}|_{\beta=\beta_H}
&=&\frac{r_h^2f^\prime(r_h)}{360}\frac{1}{\epsilon}-\left[\frac{r_h^2}{360}\left(\frac{2f^\prime (r_h)}{r_h}-f^{\prime\prime}(r_h)\right)-\frac{m^2_\phi r_h^2}{12}\right]\log \epsilon+\mathcal{O}(1)
\eea

The entropy can be expressed in terms of the proper distance of the brick wall from horizon \cite{hooft},
\bea
h_c=\int_{r_h}^{r_h+\epsilon} f^{-1/2}(r)\;dr \approx 2\sqrt{\frac{\epsilon}{f{^\prime}(r_h)}}+\mathcal{O}(\epsilon^{3/2}),
\eea
and can be rewritten  as
\bea
S_{BW}&\approx& \frac{r_h^2}{90h_c^2}-\left[\frac{r_h^2}{180}\left(\frac{15r_h^2}{(r_h^2+g^2)^{2}}-\frac{9}{(r_h^2+g^2)}-\frac{2}{r_h^2}\right)-\frac{m^2_\phi r_h^2}{6}\right]\log h_c.
\eea
The divergent terms in the entropy (in the limit uv cut-off $h_c$ approaches zero), can be absorbed in the renormalization of the coupling constants in the one-loop gravitational action \cite{nd,demers,solodukin,wald,fur,shim,wint} and one obtains the area law.

 \vspace{1cm}
\noindent
{\bf{Acknowledgements:}}
DVS would like to thank the hospitility of Dyal Singh college, University of Delhi, where the first version of this preprint was prepared.
\vspace{1cm}



\begin{thebibliography}{99}
\bibitem{JD}
J.D. Bekenstein, Phy. Rev. D {\bf 7}, 2333 (1973); Lett. al Nuovo Ciemnto 4 (1972) 15; Phy. Rev. D {\bf 9}, 3292 (1974); Phy. Rev. D {\bf 12}, 3077 (1975); Phy. Rev. D {\bf 7}, 2333 (1973).
\bibitem{fr1}
S. Fernando, Int. J. Mod. Phys. D {\bf 26}, 1750071 (2017)
\bibitem{dvs18}
S. G. Ghosh, D. V. Singh and S. D. Maharaj, Phys. Rev. D {\bf 97}, 104050 (2018).
\bibitem{ads}
A. Kumar, D. V. Singh and S. G. Ghosh Eur. Phys. Journal C 79, 275 (2019).
\bibitem{Maluf:2018lyu} 
  R.~V.~Maluf and J.~C.~S.~Neves, Phys.\ Rev.\ D {\bf 97}, 104015 (2018).
\bibitem{Ansoldi:2008jw}  S.~Ansoldi,  arXiv:0802.0330 [gr-qc].
\bibitem{javed}
W. Javed, Z. Yousaf and Z. Akhtar, Mod. Phys. lett. A \textbf{33}, 1850089 (2018).
\bibitem{Bardeen:1968}
J.~Bardeen, in {\it Proceedings of GR5} (Tiflis, U.S.S.R., 1968).
\bibitem{ABG99}E. Ayon-Beato, A. Garcia, Phys. Lett. B {\bf
 493}, 149 (2000); 
Phys.\ Rev.\ Lett.\  {\bf 80}, 5056 (1998);  Gen.\ Rel.\ Grav.\  {\bf 31}, 629 (1999); Gen. Rel. Grav.{\bf 37}, 635
(2005).	

\bibitem{Xiang} L.~Xiang, Y.~Ling and Y.~G.~Shen, Int.\ J.\ Mod.\ Phys.\ D {\bf 22}, 1342016 (2013);  H.~Culetu,  Int.\ J.\ Theor.\ Phys.\  {\bf 54},  2855 (2015);  L.~Balart and E.~C.~Vagenas,  Phys.\ Lett.\ B {\bf 730}, 14 (2014); L.~Balart and E.~C.~Vagenas,  Phys.\ Rev.\ D {\bf 90}, 124045 (2014) .

\bibitem{Singh:2017qur}
  D.~V.~Singh and N.~K.~Singh,
  Annals Phys.\  {\bf 383}, 600 (2017).
\bibitem{dym1}
I.G. Dymnikova, Int. J. of Mod. Phys. D, {\bf 5}  529 (1996); Gen. Rel. Grav. {\bf 24}, 235 (1992); Class. Quant. Grav. {\bf 21}, 4417  (2004); I. Dymnikova, M. Korpusik, Phys. Lett. B {\bf 685}, 12 (2010).

\bibitem{hamid18}
S. Hamid Mehdipour and M. H. Ahmadi, Nuc. Phys. {\bf B 926} 49 (2018).

\bibitem{singh18}
D. V. Singh, M. S. Ali and S. G. Ghosh, Int. J. Mod. Phys. D 27, 1850108 (2018).
 

\bibitem{ma14}
M. Ma and R.  Zhao, Class. Quantun Grav. {\bf{31}}  245014 (2014).
\bibitem{rashid}
D. A. Rashid ``Nonlinear electrodynamics: zeroth and first laws of black hole mechanics" arXiv: 9702087 [hep-h]. 
\bibitem{breton}
N. Breton, Gen. Rel. Grav. {\bf 37}, 643 (2005).
\bibitem{zhang}
Y. Zhang , and S. Gao, Class. Quantum Grav.{\bf 35}, 145007 (2018).
\bibitem{gulin}
L. Gulin and I. Smolic, Class. Quantum Grav.{\bf 35}, 025015 (2018).
 
 \bibitem{wald93}
R. M.  Wald, Phys. Rev. {\bf D 43}, R3427 (1993).
 

\bibitem{hooft}
`t Hooft, G, Nucl. Phys. B {\bf 256}, 727 (1985).
\bibitem{li}
Z. H. Li, Phys. Lett. B {\bf 643}, 64 (2006).
\bibitem{aghosh}
A. Ghosh, P. Mitra,  Phys. Rev. Lett. {\bf 73}, 2521 (1994); Phys. Lett. B {\bf 357}, 295 (1995).
\bibitem{eune}
M. Eune, W. Kim,  Phys. Lett. B {\bf 723}, 177 (2013).
\bibitem{sarkar}
S. Sarkar, S. Shankaranarayanan, L. Sriramkumar, Phys. Rev. D {\bf 78} 024003 (2008).
\bibitem{kghosh}
K. Ghosh, Phys. Rev. D {\bf 60} 104003 (1999).
\bibitem{solodukhin1}
S. N. Solodukhin, Phys. Rev. D {\bf 51}, 618 (1995); Phys. Rev. D {\bf 51}, 609 (1995).
 \bibitem{sharif}
M. Sharif and W Javed, J. Korean Phys. Soc. 57: 217 (2010) arxiv:1007.4995.
\bibitem{demers}
J.-G. Demers, R. Lafrance, and R. C. Myers, Phys. Rev. D {\bf 52}, 2245 ~(1995).
\bibitem{solodukin}
S. N. Solodukhin, Phys. Rev. D {\bf 54}, 3900 ~(1996).
\bibitem{nd}
N. D. Birrell and P. C. W. Davies, Quantum Fields in Curved Space ~Cambridge University Press, Cambridge, England, 1982.
\bibitem{wald}
R. M. Wald, Quantum Field Theory in Curved Spacetime and
Black Hole Thermodynamics ~University of Chicago Press, Chicago, 1994.
\bibitem{fur}
D. V. Fursaev and S. N. Solodukhin, Phys. Lett. B {\bf 365}, 51 ~1996.
\bibitem{shim}
T. Shimomura, Phys. Lett. B {\bf 480},  207 ~(2000).
\bibitem{wint}
E. Winstanley, Phys. rev D {\bf 63} 084013 (2001).
 
 



\end{thebibliography}
\end{document}